\documentclass[12pt]{article}
\usepackage{graphicx,amsmath}
 \usepackage{amssymb}
\usepackage{lineno,color}
\usepackage{url}
\begin{document}
\thispagestyle{empty}
\begin{center}
{\bf\Large Hunting for QCD Instantons at the LHC\\
 in Events with Large Rapidity Gaps}\\
 
 \vspace{1cm}
   
 V.~A.~Khoze$^{a,b}$, V.~V.~Khoze$^b$, D.~L.~Milne$^b$ and M.~G.~Ryskin$^{a,b}$\\
 
 \vspace{0.7cm}
{\small  $^a$ Petersburg Nuclear Physics Institute, NRC Kurchatov Institute, Gatchina, St.~Petersburg, 188300, Russia\\
 
 $^b$ IPPP, Department of Physics, Durham University, Durham, DH1 3LE, UK\\
}
 \vspace{0.7cm}

 \abstract{
 \noindent We outline a strategy of how to search for QCD instantons of invariant mass 20  -- 60 GeV in diffractive events in low luminosity runs at the LHC. We show that by imposing appropriate selection criteria on the final states, one can select the kinematic regime where the instanton signal exceeds the background by a factor of at least 8. In spite of the relatively strong cuts that we impose on the total transverse energy and the number of charged tracks, $\sum_i E_{T,i}>15$ GeV,  $N_{\rm ch}>20$ measured within the $0<\eta<2$ interval and excluding events with high $p_{T}$ particles, the expected cross-section is sufficiently large to study the instanton production in the events with Large Rapidity Gaps at low luminosities, thus avoiding problems with pile-up. The paper also includes an updated computation of instanton cross-sections and other parameters relevant for the ongoing studies. }   
 
 \vfill
 
 E-mail: \url{v.a.khoze@durham.ac.uk}, \url{valya.khoze@durham.ac.uk}, \\
 \quad \quad \, \url{daniel.l.milne@durham.ac.uk}, \url{ryskin@thd.pnpi.spb.ru}
 
  \end{center}
 \newpage
 
 \section{Introduction}

Instantons are non-perturbative classical solutions of Euclidean equations of motion in non-abelian gauge theories~\cite{BPST}. 
In the semi-classical limit, instantons provide dominant contributions to the path integral and describe quantum tunnelling between different vacuum sectors of the theory \cite{tH,Callan:1976je,Jackiw:1976pf}. Over the years a number of phenomenological models and approaches were developed where QCD instantons were either directly responsible for generating, or at least contributed to many key aspects of non-perturbative low-energy dynamics of strong interactions~\cite{tHooft:1986ooh,Callan:1977gz,Novikov:1981xi,Shuryak:1982dp,DP,Schafer:1996wv}.
These include
the role of instantons in the breaking of the $U(1)_A$ symmetry and the spontaneous breakdown of the chiral symmetry, the formation of quark and gluon condensates, topological susceptibility of the vacuum, and the non-perturbative generation of the axion potential. 
Instanton-based phenomenological models provide at least a qualitative description of the QCD vacuum.\footnote{Their potential short-comings originate from attempting to describe the strongly-coupled QCD vacuum dynamics using a semi-classical (weak-coupling) approximation where the path integral is expanded around classical instanton background fields.} These predictions are in good agreement with lattice calculations
\cite{Hasenfratz:1998qk}, see also \cite{DeGrand:1997gu,Smith:1998wt}.

\medskip

Our motivation here is different. We would like to search for QCD instanton events directly in scattering experiments. This setup provides a more controlled environment than QCD vacuum modelling, as the scattering event kinematics will enable us to remain in a relatively weakly-coupled regime where semiclassical instanton methods are reliable. In scattering processes the instanton can be thought of as new non-perturbative multi-particle vertices in a Feynman diagram, and for its computation at the microscopic level we will use the formalism developed in the recent papers~\cite{KKS,KMS}.
QCD instanton-generated processes are predicted to be produced with a large scattering
cross-section at small centre-of-mass partonic energies, although, as shown in \cite{KMS}, discovering them at hadron colliders using conventional high-$p_T$ trigger requirements, is a challenging task that requires alternative dedicated search strategies. Developing such a strategy based on the event selection with large rapidity gaps at low luminosities is the purpose of this paper.  

\medskip

In the past, in the context of deep inelastic scattering,  QCD instanton contributions~\cite{Balitsky:1993jd,Moch:1996bs,Ringwald:1998ek} were searched for (but not observed)~\cite{Adloff:2002ph, Chekanov:2003ww} at the HERA collider.

\medskip

In QCD, the instanton configuration consists of the gauge field,
\begin{eqnarray}
A_\mu^{a\, {\rm inst}}(x) &=& \frac{2 \rho^2}{g} \frac{\bar{\eta}^a_{\mu\nu} (x-x_0)_\nu}{(x-x_0)^2((x-x_0)^2+\rho^2)}\,,
\label{eq:instFT1}
\end{eqnarray} 
along with the fermion components for light ($m_f <1/\rho$) fermions,
\begin{equation}
\bar{q}_{Lf} = \psi^{(0)}(x) \,, \quad q_{Rf} = \psi^{(0)}(x) \,.
\label{EQ_i_gauge}
\end{equation}
The gauge field $A_\mu^{a\, {\rm inst}}$ is the Belavin-Polyakov-Schwartz-Tyupkin (BPST) instanton solution~\cite{BPST}  of the self-duality equations in the singular gauge. Here $\rho$ is the instanton size and $x_0$ is the instanton position. 
Constant group-theoretic coefficients $\bar{\eta}^a_{\mu\nu}$ are the 't Hooft eta symbols defined in \cite{tH}. The fermionic components $\psi^{(0)}$ are the corresponding normalised solutions of the Dirac equation
 $\gamma^\mu D_\mu[A_\mu^{a\, {\rm inst}}] \psi^{(0)} \,=0$. These are the fermion zero modes of the instanton.
The instanton configuration is a local minimum of the Euclidean action, and the action on the instanton is given by
$S_I= \frac{8\pi^2}{g^2} = \frac{2\pi}{\alpha_s}$.

The instanton configuration  \eqref{eq:instFT1} has topological charge equal to one and thus, due to the chiral anomaly, the instanton processes violate chirality. If the instanton is produced by a two-gluon initial state, the final state of this instanton-mediated process will have $N_f$ pairs of quarks and anti-quarks with the same chirality,
\begin{equation}
\label{e2}
g+g\to n_g\times g + \sum^{N_f}_{f=1}(q_{Rf}+\bar q_{Lf})\ ,
\end{equation} 
where $N_f$ is the number of light flavours relative to the inverse instanton size, $m_f <1/\rho$.
The instanton contribution to the amplitude for this process comes from expanding the corresponding path integral in the instanton field background.
At leading order in the instanton perturbation theory, the amplitude
takes the form of an integral over the instanton collective coordinates (see e.g. Refs.~\cite{KKS,KMS} for more detail),
\begin{equation} 
{\cal A}^{\rm \, L.O.}_{\, 2\to\, n_g+ 2N_f} = \int d^4 x_0 \int_0^\infty d\rho \, D(\rho) \, e^{-S_I}\,
\prod_{i=1}^{n_g+2} A_{{\rm LSZ}}^{\rm inst}(p_i;\rho)\, \prod_{j=1}^{2N_f} \psi^{(0)}_{{\rm LSZ}}(p_j; \rho).
\label{EQ_ampgg}
\end{equation}
The factors of $A_{{\rm LSZ}}^{\rm inst}(p_i;\rho)$ and 
$ \psi^{(0)}_{{\rm LSZ}}(p_j; \rho)$ are the standard insertions of the LSZ-reduced instanton fields in the momentum representation and $D(\rho)$ is given by the known expression for the instanton density~\cite{tH}.
\medskip

From the point of view of
Feynman graphs the leading order instanton amplitude \eqref{EQ_ampgg} reveals itself as a family of multi-particle vertices (with different numbers of emitted gluons) integrated over the instanton position and size. It describes the
emission of a large number of gluons, $n_g \propto E^2/\alpha_s$, together with a fixed number of quarks and anti-quarks, one pair for each light flavour in accordance with \eqref{e2}.
The semi-classical suppression factor, $\exp(-S_I)=\exp(-2\pi/\alpha_s)$, will be partially compensated by the growth with jet energy, $E$, of the high multiplicity cross-section for the process  \eqref{e2}.
The fully factorised structure of the field insertions on the right hand side of \eqref{EQ_ampgg} implies that at leading order in instanton perturbation theory there are no correlations between the momenta of the external legs in the instanton amplitude. The momenta of individual particles in the final state are mutually independent, apart from overall momentum conservation.

\medskip

Thus to discover the QCD instanton we have to observe in the final state a multi-particle cluster or a fireball which contains in general a large number of isotropically distributed gluon (mini)jets accompanied by $N_f$ pairs of light quark jets generated by a subprocess such as \eqref{e2}. 

\medskip

 It is quite challenging, however, to identify the instanton on  top of the underlying event. Recall that the instanton is not a particle and there will be no peak in the invariant mass, $M_{\rm inst}$, distribution~\footnote{What we mean by the instanton mass is the partonic energy $\sqrt{\hat{s}}$ of the initial 2-gluon state in the process \eqref{e2}. As we integrate over the Bjorken x variables when computing hadronic cross-section, we sum over a broad range of instanton masses. }.
 The mean value of $M_{\rm inst}$ can at least in principle be ``measured" or reconstructed as the mass of the minjet system created by the instanton fireball in each given event. Talking about the instanton we actually mean a family of objects of different sizes, $\rho$ and different orientation in the colour and the Lorentz spaces. The mean value of $M_{\rm inst}$ depends on $\rho$, increasing when $\rho$ decreases. Since experimentally it is impossible to measure the instanton size, $\rho$, below we  use the  mass $M_{\rm inst}$ to characterise the properties of the instanton production event.
 
 \medskip
 
The paper is organised as follows; in Section 2 we detail the main sources of background and how their cross-section behaves with energy. Section 3 details the physics of events with large rapidity gaps while Section 4 explains the rationale behind the selection criteria employed in this paper.  Section 5 updates the calculation of cross-sections presented in \cite{KMS} to account for small virtualities present in incoming gluons
and provides details of the Monte Carlo generation of our background. In Section 6 we give our results and Section 7 provides a discussion theoretical uncertainties and additional instanton sub-processes. In section 8  we present our conclusions.

\section{ Background }

Let us consider the main sources of background which can mimic the instanton signal.

\medskip

The cross-section of instanton production falls steeply with $M_{\rm inst}$ mainly due to the factor $\exp(-S_I)=\exp(-2\pi/\alpha_s(\rho))$ in the amplitude. At  smaller values of $\rho$ the instanton action
$S_I=2\pi/\alpha_s(\rho)$ increases since the QCD coupling $\alpha_s$ decreases. Calculations presented in Section 5 ({\it cf.}~Table~\ref{Tab_1}) show that the elementary cross-section of the parton level subprocess (\ref{e2}) falls approximately as  
\begin{equation}
\hat\sigma_{\rm inst}\propto M^{-6}_{\rm inst}\,,
\label{e_inst_simp}
\end{equation}
over a broad low-to-intermediate energy range
(it becomes less steep, $\hat\sigma_{\rm inst}\propto M^{-4}_{\rm inst} $, at lower energies 20 -- 30 GeV).

\medskip

On the other hand the dimensionless rate, $M^2 \, \sigma_{\rm pQCD}$, of a similar purely perturbative QCD subprocess, 
$gg\to N{\rm minijets}$, decreases only logarithmically. 
For the perturbatively formed `hedgehog' configuration of $N$ final state jets we would expect
\begin{equation}
\label{e3}
\sigma_{\rm pQCD}(gg\to N\,{\rm jets})\, \sim\,  \frac{16\pi}{M^2}\left(\frac{N_c}\pi\, \alpha_s(M)\right)^N\,,
\end{equation}
where $M$ denotes the invariant energy of the perturbatively formed cluster of minijets.
Thus, at sufficiently large values of $M_{\rm inst}$ the instanton signal~\eqref{e_inst_simp} will become negligible relative to the purely perturbative QCD production~\eqref{e3}.

\medskip

\begin{figure} [t]
\begin{center}
\includegraphics[trim=0.0cm 0cm 0cm 0cm,scale=0.47]{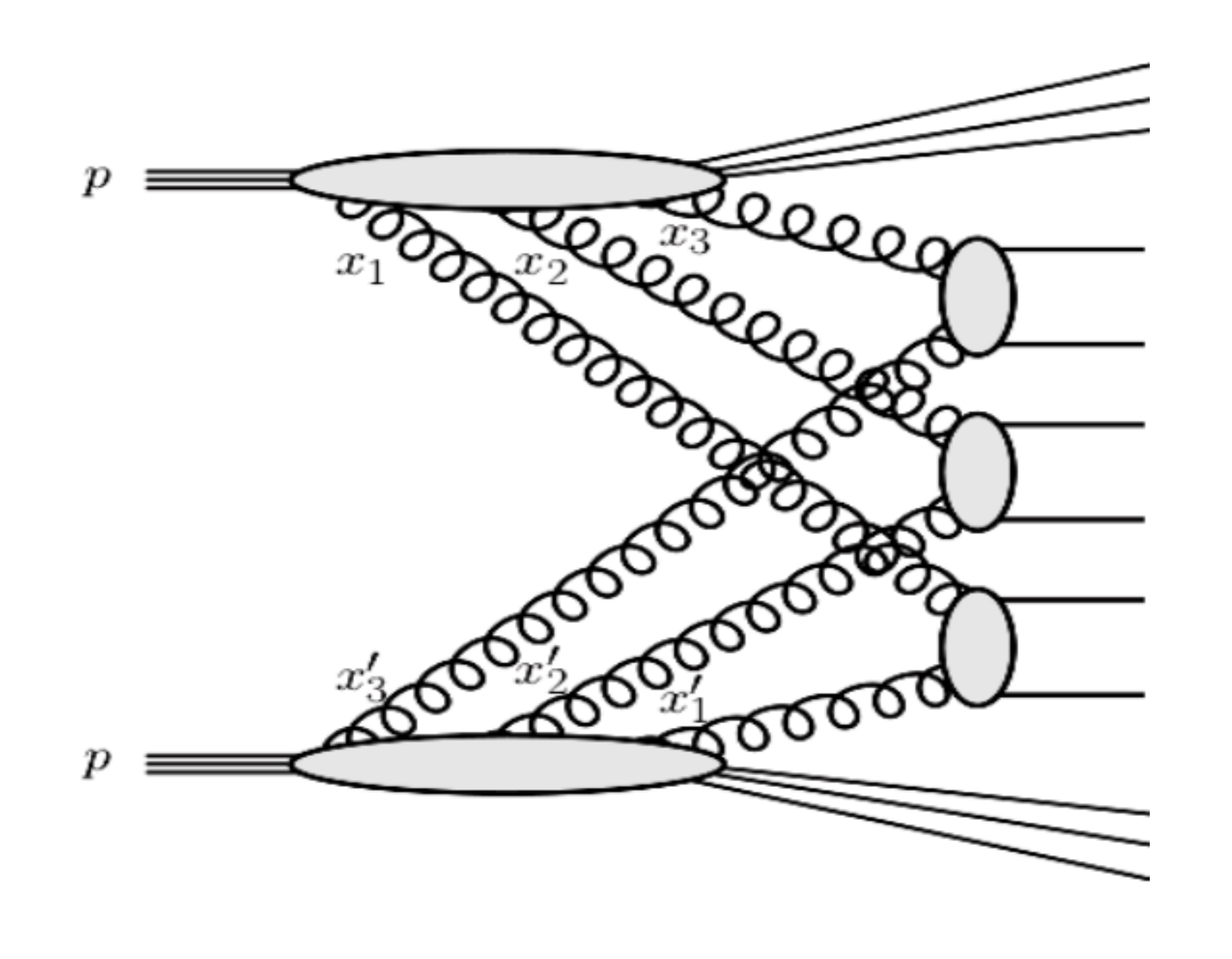}
\caption{\small Schematic diagram for the triple parton scattering process.} 
\label{f1}
\end{center}
\end{figure}

In the regime of interest of moderately small $M_{\rm inst}$ we face however another problem that needs to be addressed.
The instanton event can be mimicked by the multiple parton interactions (MPI) illustrated in Fig.~\ref{f1}. Indeed, the double (triple, ..., $n$) parton scattering produces a few pairs of jets (dijets) which would  look like a fireball, thus obscuring the final state signature of the genuine instanton signal.

The cross-section of such a process can be evaluated as
\begin{equation}
\frac{d\sigma}{dE^2_1...dE^2_n}\, \sim\, \left(\frac{d\sigma}{\sigma_{eff}dE^2_1}\ ...\ \frac{d\sigma}{\sigma_{eff}dE^2_n}\right)\sigma_{eff}\ ,
\end{equation}
where  $E_i$ denotes the transverse energy of a jet in the $i$ dijet system, 
$\sigma_{eff}\sim 10$~mb (see Fig.~4 of~\cite{Aaboud:2018tiq} and references therein) and the cross
section $d\sigma/dE^2_i\sim \pi\alpha_s^2/E^4_i$.
That is, at low $M_{\rm inst}$ when the transverse energies of produced jets become small, the MPI processes will dominate.
It was shown in~\cite{Sas:2021yxx} that the events caused by such multiple parton interactions manifest high sphericity S, similar to the instanton signal. Our strategy to suppress them is outlined in the following section and involves a final state selection with Large Rapidity Gaps.

\medskip

Finally, we would like to emphasise one additional point.
Since our knowledge on the dynamics of confinement and hadronization
is quite limited,
general purpose Monte Carlo (MC) event generators 
introduce  
certain phenomenological parameters that are tuned in order to better describe the experimental data (for a  review see e.g. Ref.~\cite{Buckley:2011ms}).
 But  if a low-mass  instanton contributes to the scattering, its presence will affect the data, and after
such  tuning we lose the possibility to distinguish the low $M_{\rm inst}$ instanton
from other hadronization effects.

\medskip
\section{Events with Large Rapidity Gaps}
\medskip

As discussed above, we expect a large `underlying event' background both in the region of  large and  small instanton masses. 
However the low $M_{\rm inst}$ background caused by multi-parton interactions can be effectively suppressed by selecting  events with  Large Rapidity Gaps (LRG). Indeed, each additional 'parton-parton$\to$ dijet' scattering is accompanied by the colour flow created by the parton cascade needed to form the incoming partons (see Fig.~\ref{f1}). This colour flow produces secondaries which fill the LRG. The LRG survival probability, $S^2$, (i.e. the probability not to destroy the LRG) is rather small, $S^2\leq 0.1$, see e.g.~\cite{LRG}.
 Thus the probability to observe $n$ additional branches of parton-parton interactions in LRG events is suppressed by the factor $(S^2)^n$. 

\medskip

 Moreover, recall that the events with an LRG mainly occur at large values of $b_t$, (the separation between the two incoming protons in the transverse plane), where the
  proton optical density (opacity) is small~\cite{LRG,SD}.
 On the other hand, the MPI (Double/Triple/.../n Parton Scattering) processes  proceed dominantly at small values of  $b_t$, where the proton opacity is much larger. Thus, we expect that the suppression factor, $S^2$, for each additional scattering should be smaller than the average value of
  $S^2$ discussed in~\cite{LRG}. Therefore in this paper when  evaluating  backgrounds for relatively low mass instantons in the events with an LRG, to a first approximation we neglect the MPI/nPS contributions.

  \section {Search strategy}
 \medskip

We propose searching for the instanton as a multi-particle cluster/fireball with a mass of about 20 -- 60 GeV in the events with an LRG. The presence of an LRG can be detected either by detecting the leading forward  proton with beam momentum fraction, $x_L=1-\xi$, very close to 1 ($\xi=x_{Pom}\leq 0.01$), or by observing no hadron activity in the forward calorimeters.

\medskip

Since the relatively heavy instanton produces  a rather large number of jets we are looking for 
high multiplicity events which:
\begin{itemize}
\item do not contain high-$E_T$ jets, and 
\item still have a large density of the transverse energy, $\sum_i dE_{Ti}/d\eta\sim M_{\rm inst}/3$ (the sum is over all  secondary particles in the given $\eta$ interval).
\end{itemize}
Indeed, from Table~\ref{Tab_1} we see that the instanton of mass 30 GeV  produces about 17 jets (9 gluons plus 4 light $\bar qq$ pairs).
The energy of each jet $E_{Ti}\sim 1/\rho\sim 2$ GeV.
After hadronization in such an event we expect about 40-60 particles. The large multiplicity can be used as the main (or additional) trigger to select the events of interest.
 
 \medskip
 
Note that the elementary (parton level) instanton cross-section is predicted to be rather large ($\sim 1 \mu b$) while the  background due to hedgehog-like pQCD multijet production (\ref{e3}) is much smaller -- already for the case of  
6 final jets the expected elementary (hedgehog QCD) cross-section $\sim \mbox{few}\cdot$nb which can be safely neglected.  

\medskip

On the other hand, in the events with an LRG, the probability to observe such large multiplicity in a limited rapidity interval falls steeply with $N_{track}$ (see e.g. Fig.~13 of~\cite{SD}). Moreover, since each jet from the instanton cluster
  contains a leading hadron with $p_T\geq 1$ GeV, we can select the events with a large multiplicity (say, $N_{ch}(p_T>0.5\mbox{GeV})>20$) of high $p_T$ particles, in this way strongly suppressing the soft QCD background.
  
\begin{figure} [t]
\begin{center}
\includegraphics[trim=0.0cm 0cm 0cm 0cm,scale=0.47]{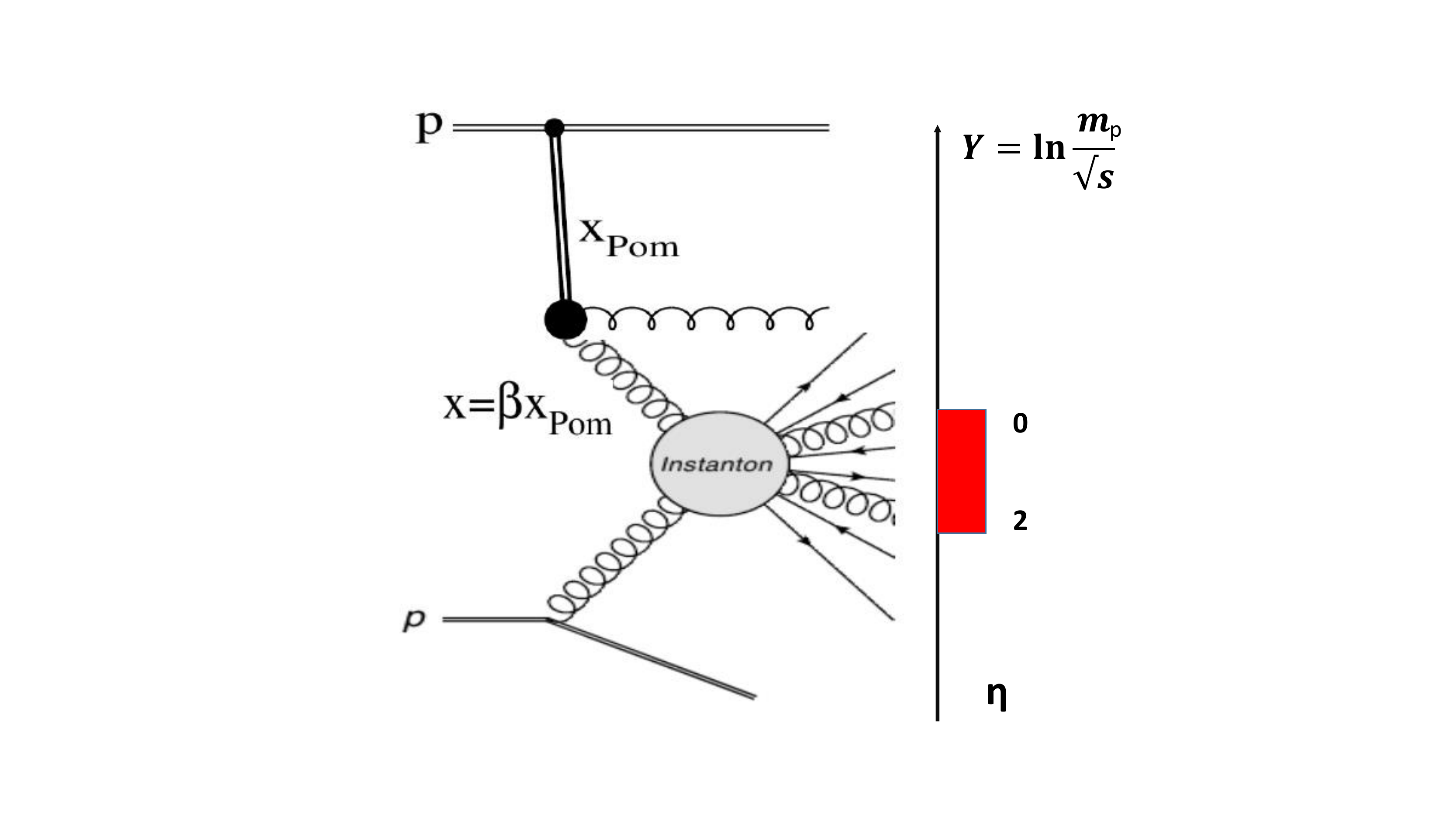}
\caption{\small Instanton production in a diffractive process with an LRG. The pomeron exchange is shown by the thick doubled line. The red bar shows the range of $\eta$ considered in this paper. $Y$ indicates the incoming proton position in rapidity. As shown in the diagram secondaries will be produced outside this range but will not be used when calculating $E_{T}$ or $N_{ch}$.}  
\label{f2}
\end{center}
\end{figure}

\medskip

  Since the LHC detectors never cover the whole 
  ($4\pi$) rapidity interval, there is no chance to adequately measure the value of $M_{\rm inst}$. To select the events with appropriate $M_{\rm inst}$ we introduce the cut on the total transverse energy measured within the given rapidity interval $\sum_i E_{Ti}>M_0$.\\
  
  Here we consider the instanton production in the pomeron-proton  collision (in terms of Regge theory  the pomeron exchange is responsible for the presence of the LRG) selecting  events with a large multiplicity and relatively large transverse energy~\footnote{The idea to search for the instanton in events with very large multiplicity but not too large transverse energy and the first evaluation of the instanton cross-section at collider energies was discussed long ago in~\cite{BR}.}.
We expect that these events will be more or less spherically symmetric, that is, in such events there should be a large probability to observe the sphericity $S$ close to 1. Since we can not observe the particles in the whole $4\pi$ sphere we consider the ``transverse sphericity" (or  cylindricity) defined as $S_T=2\lambda_2/(\lambda_1+\lambda_2)$ where $\lambda_i$ are the eigenvalues of the matrix
  \begin{equation}
  \label{st1}
  S^{\alpha\beta}=\frac{\sum_i p^\alpha_i p^\beta_i}{\sum_i |\vec p^2_i|}\ ,
  \end{equation}
  and $\lambda_2<\lambda_1$.
Here $p_i^\alpha$ is the two-dimensional transverse component of the momentum of the $ i$-th particle and we sum over all particles observed in the event within a given rapidity interval.\\

Finally recall that the decay of the instanton produces one additional pair of each flavour of light ($m_f<1/\rho$) quark. So in the case of the signal we expect to observe a  larger number of strange
and charm particles than in background events.

\section{Calculation}

\subsection{Parton-level instanton cross-sections}

The instanton cross-sections of the elementary gluon-initiated process \eqref{e2} are calculated following the approach developed in the recent papers~\cite{KKS,KMS}.
In this approach the total parton-level instanton cross-section for the process $gg \to X$ is related by the optical theorem to the imaginary part of the
forward elastic scattering amplitude computed in the background of the instanton--anti-instanton ($I\bar{I}$) configuration,
\begin{eqnarray}
\hat\sigma_{\rm inst} &=&  \frac{1}{E^2}\, {\rm Im} \, {\cal A}^{{I\bar{I}}}_4 (p_1,p_2,-p_1,-p_2) 
\,.
\label{EQ_opt1}
\end{eqnarray}
Here $E= \sqrt{(p_1+p_2)^2}$ denotes the invariant mass $\sqrt{\hat{s}}$ of the 2-gluon initial state. The resulting instanton cross-section takes the form of the finite-dimensional integral~\cite{KMS} over the instanton and anti-instanton collective coordinates (scale-sizes, $\rho$ and $\bar{\rho}$, the separation between the instanton and anti-instanton centres, $R$, and the relative $I\bar{I}$ orientations, $\Omega$):
 \begin{eqnarray}
\label{eq:firstint}
&&\hat\sigma_{\rm tot}^{\rm inst} \,\simeq \,
\frac{1}{E^2}\,{\rm Im}\, \frac{\kappa^2 \pi^4}{36\cdot4}   \int \frac{d\rho}{\rho^5}
 \int \frac{d\bar{\rho}}{\bar{\rho}^5}  \int d^4 R \int d\Omega  \left( \frac{2\pi}{\alpha_s(\mu_r)}\right)^{14} (\rho^2 E)^2 (\bar{\rho}^2E)^2\,
  {\cal K}_{\rm ferm}(z)
\nonumber
  \\
&& (\rho\bar\rho \mu_r^2)^{b_0} \,\exp\left(R_0 E\,-\,\frac{4\pi}{\alpha_s(\mu_r)} \,{\cal S}(z)
\,-\, \frac{\alpha_s(\mu_r)}{16\pi}(\rho^2+\bar{\rho}^2) \, E^2\, \log \frac{E^2}{\mu_r^2}\,-\, Q(\rho+\bar\rho)
\right).\nonumber\\
 \label{EQ_opt2}
\end{eqnarray}
In the above integral $\kappa^2 $ denotes the known normalization constant and ${\cal K}_{\rm ferm}(z)$ is the factor arising from the overlap of fermion zero modes of $I$ and $\bar{I}$ and the variable $z$ is a certain conformally-invariant ratio of the collective coordinates $R$, $\rho$ and $\bar\rho$~~\cite{Yung:1987zp,Khoze:1991mx},
\begin{equation}
z\,=\, 
\frac{R^2+\rho^2+\bar{\rho}^2+\sqrt{(R^2+\rho^2+\bar{\rho}^2)^2-4\rho^2\bar{\rho}^2}}{2\rho\bar{\rho}}\,.
\label{EQ_defz}
\end{equation}
The first factor in the exponent in \eqref{EQ_opt2} describes the energy input from the initial state, the second factor is the action of the instanton-anti-instanton configuration~\cite{Khoze:1991mx}, $S_{I\bar{I}}(z)= \frac{4\pi}{\alpha_s} \, {\cal S}(z)$ where,
\begin{eqnarray}
{\cal S}(z) \,=\, 3\frac{6z^2-14}{(z-1/z)^2}\,-\, 17\,-\, 
3 \log(z) \left( \frac{(z-5/z)(z+1/z)^2}{(z-1/z)^3}-1\right)\,.
 \label{EQ_defSz}
\end{eqnarray}
The third term in the exponent,
\begin{equation}
\exp\left(-\, \frac{\alpha_s(\mu_r)}{16\pi}\,(\rho^2 +\bar\rho^2)\, E^2\, \log \frac{E^2}{\mu_r^2}\right)\,,
\label{EQ_mff}
\end{equation} 
 is the quantum effect that takes into account resummed perturbative exchanges between the hard initial state gluons and the instantons~\cite{Mueller:1990qa}.
Inclusion of these quantum effects is required in order to resolve the well-known infra-red problem in the $\rho,\bar{\rho}\to \infty$ limit, as QCD instantons with $\rho \gg \left( \frac{16\pi}{\alpha_s}\frac{1}{E^2 \log E^2}\right)^{1/2}$
are automatically cut-off by these quantum corrections.

\medskip

We note that the initial-state partons (in our case gluons) are not strictly on mass-shell, but  carry small order-GeV virtualities, $Q_1$ and $Q_2$. For the gluon emitted from the pomeron we choose $Q_1 = 2$~GeV, and the second gluon that originated from the proton has the virtuality $Q_2 = 1$~GeV. \footnote{From the pomeron side the expected  virtuality $Q^2_1\simeq q^2_t$ is close to the mean transverse momentum squared, $q^2_t$, of the gluon inside the pomeron. Its  value can be estimated as $q^2_t\sim 1/\alpha'_P\sim 4$ GeV$^2$ where we take the pomeron trajectory slope $\alpha'_p=0.25$ GeV$^{-2}$~\cite{Donnachie:1983hf}. The gluon virtuality in the proton PDF, $Q_2\sim 1$~GeV was chosen for the following reasons. As a rule the global parton analysis based on DGLAP evolution starts the evolution at some $Q=Q_0\geq 1$ GeV. We do not know the parton distributions at a smaller $Q$ (it is not even clear whether we can use the parton language at such small $Q$). Note however that the low-$x$ gluon density increases as the scale $Q$ increases. So we consider the value of $Q_2=1$ GeV as a conservative estimate.} 
These virtualities  introduce the 
form-factor\footnote{The form-factor is a direct consequence of Fourier transforming the instanton field 
to momentum space to obtain $A^{\rm inst}_{LSZ}(p_i)$, where the momentum $p_i$ has the virtuality, $Q_i^2$~\cite{Khoze:1991mx}, and was explored and used extensively in the context of deep inelastic scattering in 
 \cite{Balitsky:1993jd,Ringwald:1998ek}.} $e^{-Q\rho}$ in the instanton vertex, with $Q=Q_1+Q_2=3$~GeV in our case. This is the origin of the final term, 
 \begin{equation}
\exp\left(-Q (\rho+\bar{\rho})\right)\,,
\label{EQ_qqvirt}
\end{equation}
 in the exponent of the cross-section in \eqref{EQ_opt2}.
  
 \medskip
 
\noindent To further simplify the integrand we select a natural value for the renormalization scale dictated by the inverse instanton size.
This prescription removes the large $ (\rho \mu_r)^{b_0} (\bar{\rho} \mu_r)^{b_0}$ factor in front of the exponent in \eqref{EQ_opt2}.
Hence we choose,
\begin{equation}
\mu_r \,=\, 1/\langle \rho\rangle \,=\, 1/\sqrt{\rho \bar{\rho}}\,.
\label{eq:mdef}
 \end{equation} 
 
 For the reference point of $\alpha_s$ we choose its  value at the $\tau$ mass, 
 $\alpha_s(m_\tau)= 0.32$ following~\cite{Beneke:2008ad,Abbas:2012fi} that fits the experimental data~\cite{Schael:2005am, Ackerstaff:1998yj}.
 The choice of $m_\tau$ as the reference scale for $\alpha_s$ is conveniently close to the interesting (for us) regime of relatively light instantons with  sizes  $1/\langle \rho\rangle \sim$~few~GeV.
 For the running coupling at energy scales above $m_\tau$, we use the 1-loop expression, while for the lower scales we freeze the coupling at a fixed critical value $\alpha_s= 0.35$,
 \begin{equation}
 \frac{4\pi}{\alpha_s(\langle\rho\rangle)}    \, \simeq \, 
\begin{cases}
\quad     \frac{4\pi}{0.32}    \,-\, 2 b_0 \log\left( \langle \rho\rangle\,m_\tau\right)  & :\,\,{\rm for}\,\,  \langle\rho\rangle^{-1} \ge1.45\,{\rm GeV}\\
\quad \frac{4\pi}{0.35}  & :\,\,{\rm for}\,\, \langle\rho\rangle^{-1} <1.45\,{\rm GeV}\, .
\end{cases}
\label{eq:alphadef}
\end{equation}
At these energy-scales we are in the regime of 
\begin{equation}
N_f=4\,,
\end{equation}
active quarks, and this is the $N_f$ value we use in $b_0=11-2 N_f/3 $ and in the instanton density expressions for $\kappa^2$ and ${\cal K}_{\rm ferm}$ in \eqref{EQ_opt2}.

The rationale for freezing the coupling at 0.35 is that the perturbative evolution of strong coupling at the order $\sim 1$~GeV scale is known to be in conflict with observations. The strong coupling values extracted from the fits to charmonium spectrum give  $\alpha_s\simeq  0.35$
at 1 to 1.2~GeV scale \cite{Badalian:1999fq}, which is close to the measured value $\alpha_s(m_\tau)$ at the considerably higher scale
 $m_\tau= 1.777$~GeV, and  is about 50\% below the prediction of the 2-loop perturbative running. These considerations justify freezing the strong coupling in the infrared at a critical value $\alpha_s=0.35$, the approach we will follow in accordance with \eqref{eq:alphadef}. 
 
\begin{table}[]
\centering
\scalebox{0.9}{
\begin{tabular}{|l|l|l|l|l|}
\hline
E [GeV]& $\hat{\sigma}\left(gg\rightarrow I\right)$ [pb] & $\langle1/\rho\rangle$ [GeV]& $\alpha_{s}\left(1/\rho\right)$ & $\langle n_{g} \rangle$ \\
\hline
20  & $2.01\times10^6$  & 1.69 & 0.327 & 7.81  \\
25  & $9.49\times10^5$  & 1.98 & 0.306 & 8.58  \\
30  & $4.64\times10^5 $  & 2.27 & 0.290 & 9.07  \\
35  & $2.32\times10^5$  & 2.52 & 0.279 & 9.61  \\
40  & $1.25\times10^5$  & 2.84 & 0.267 & 9.67  \\
50  & $3.89\times10^4$  & 3.38 & 0.251 & 10.56 \\
60  & $1.38\times10^4$  & 3.87 & 0.241 & 10.89 \\
70  & $5.45\times10^3$  & 4.33 & 0.232 & 11.38 \\
80  & $2.36\times10^3 $ & 4.85 & 0.224 & 11.67 \\
90  & $1.08\times10^3$ & 5.24 & 0.219 & 12.31  \\
100 & $5.44\times10^2$  & 5.82 & 0.213 & 12.10  \\
110 & $2.79\times10^2$  & 6.21 & 0.209 & 12.62 \\
120 & $1.53\times10^2$  & 6.71 & 0.205 & 12.77 \\
130 & $8.56\times10^1$  & 7.13 & 0.201 & 13.04 \\
140 & $4.99\times10^1$  & 7.57 & 0.198 & 13.25 \\
150 & $3.01\times10^1$  & 8.00 & 0.195 & 13.45 \\
\hline
\end{tabular}
}
\caption{\small The instanton production cross-section for the sub-process \eqref{e2} for a range of partonic cms energies $E$ (i.e. instanton masses) and the mean number of emitted gluons in addition to the mean value of $1/\rho$ and $\alpha_{s}\left(1/\rho\right)$. For energies larger than 150 GeV, the suppression coming from the 3 GeV virtuality is fairly negligible and one can use the numbers given in \cite{KMS}. Note that the mean value of $R/\rho$ was $\sim$1.55 at all energies.}
\label{Tab_1}
\end{table}

\medskip
The authors of Ref.~\cite{KMS}, evaluated the integral in \eqref{EQ_opt2} numerically using the python package SciPy \cite{Virtanen:2019joe} to derive the parton-level instanton cross-sections along with the mean number of gluons in the final state as functions of partonic energy $E$. In this work we follow the computational approach of~\cite{KMS}, except that we now also account for the effect of $Q\simeq 3$~{\rm GeV} gluon virtualities 
\eqref{EQ_qqvirt} from the outset, and use the prescription \eqref{eq:alphadef}  for the running coupling.

\medskip

The data in Table~\ref{Tab_1} shows that the instanton production cross-section $\hat\sigma_{\rm inst}$ is sharply suppressed at high partonic energies (high instanton masses) and becomes large at low values of $E$. This is also consistent with the earlier calculations in \cite{KKS}  that used the same theory set-up for the parton-level instanton cross-section \eqref{EQ_opt2},
but evaluated the integral in the steepest descent approximation. The sharp suppression of the instanton cross-section at large partonic energies is the consequence of the inclusion of the quantum exchanges between the hard initial gluons that resulted in the exponential  form-factor 
\eqref{EQ_mff}. It cuts off the (anti)-instanton sizes more and more efficiently when $E$ increases, ultimately leading to the integral being dominated by smaller and smaller instantons, thus exponentially suppressing the overall cross-section, in agreement with the results shown in Table.~1 and Fig.~6 in Ref.~\cite{KKS}. In the low-energy limit, contributions of large instantons with $\rho \gg 1/ (3{\rm GeV})$ are cut-off by gluon virtualities in \eqref{EQ_qqvirt}.

\medskip

For more detail on the theory formalism leading to \eqref{EQ_opt2} and the resulting  evaluation of $\hat\sigma_{\rm inst} (E)$ and $\langle n_g\rangle (E)$ we refer the reader to Ref.~\cite{KMS}.

\medskip
\subsection{Hadronic cross-sections for signal and background}

\medskip

The cross-section of instanton production in proton-proton collisions reads,
\begin{equation}
\sigma_{pp\rightarrow I} \,=\, \int_{\hat{s}_{min}}^{s_{pp}} dx_{1}dx_{2}\quad f\left(x_{1},Q_1^{2}\right)f\left(x_{2},Q_2^{2}\right) \hat\sigma_{\rm inst}\left(E^2=x_{1}x_{2}s_{pp}\right)\,,
\label{EQ_i_hadr_gg}
\end{equation}
where ${s}_{pp}$ is the centre-of-mass energy of the hadron collider and $\hat{\sigma}$ is the partonic instanton cross-section. Selecting the events with an LRG or the leading proton we have to replace one parton distribution (PDF), say $f(x_1,Q_1^2)$, which describes the probability to find an appropriate parton (gluon) in the incoming proton, by the so-called diffractive PDF, $f^D(x_1,Q_1^2)$, which describes the corresponding  distribution in the pomeron.

The lower limit  $\hat{s}_{min}$ in the integral \eqref{EQ_i_hadr_gg} is introduced for technical reasons. Strictly speaking, one is supposed to integrate  from $\hat{s}_{min}=0$ to ${s}_{pp}$.  However instantons of masses much smaller than 20 GeV are not under theoretical control, the semiclassical dilute instanton approximation becomes invalid for large instantons with $\rho \sim$~GeV, where the theory can no longer be considered weakly-coupled.\footnote{Furthermore, when
$\rho \gtrsim 0.3 \, {\rm fm}\sim 1.5 \,{\rm GeV}^{-1}$ one can no longer distinguish between the instantons created in collisions and the instanton-like configurations populating the QCD ground state~\cite{Hasenfratz:1998qk}. The role of $\hat{s}_{min}$ is to cut-off such large ground-state field configurations.}
It is impossible to fix $\hat{s}_{min}$ experimentally, instead we have to choose and impose
experimental cuts on final states that will suppress the contribution of the low mass instantons. In practice we will set 
$\sqrt{\hat{s}_{min}}=20 \,\text{GeV}$ in the integral,\footnote{Taking $\hat{s}_{min}=\left(20 \text{GeV}\right)^{2}$ gives a hadronic instanton cross-section of ~1.72 $\mu$b (after multiplying by $S^{2}$ and taking the factorisation scale as $1/\langle\rho\rangle$).} and then impose the cuts.

\medskip

Instanton samples are generated using the RAMBO algorithm \cite{Kleiss:1985gy} and then showered and hadronised using PYTHIA 8 \cite{Sjostrand:2014zea}. The scale of the process was taken to be $1/\langle\rho\rangle$ for the purposes of showering in PYTHIA. 
As already mentioned, in the integral \eqref{EQ_i_hadr_gg} we set the minimum instanton mass generated to be $\hat{s}_{min}=20$~GeV. 
However we also note that the inclusion of these low mass instantons is expected to give a negligible contribution with our choice of selection criteria.

\medskip

Background events were simulated by generating inelastic pion-proton collisions (see Fig.~\ref{f2}) in PYTHIA for the proton energy $E_p=7$ TeV and the pomeron energy $E_{Pom}=7\cdot x_{Pom}$ TeV (with a pion being used to simulate a pomeron) and $x_{Pom}$ fixed at 0.003. For the proton PDF we use the NNPDF3.1luxQED NNLO \cite{Buckley:2014ana,Bertone:2017bme,Manohar:2017eqh,Manohar:2016nzj} set while for the parton distribution in the pomeron (mimicked in PYTHIA by the pion) the H1 fit B~\cite{H1} is used. The $Q^2$ evolution for the pomeron PDF is performed by QCDNUM~\cite{QCDNUM}. 

\medskip

When computing the signal, the instanton cross-section calculated in \eqref{EQ_i_hadr_gg} was multipled by the gap survival probabilty, $S^{2}$. For simplicity we took $S^2=0.1$ in agreement with the experimental results~\cite{Ddijet,Ddijet2} for diffractive high $E_T$ dijet  production.
The background cross-section was normalised to the experimental yield  of single dissociation  $d\sigma^{SD}/d\ln(1/x_{Pom})\simeq 0.15$ mb~\cite{ATL}.
In generating both the instanton and the background samples, multi-parton interactions were disabled in PYTHIA.

\medskip

Since in the case of soft QCD background only a very small fraction of events satisfy the $\sum_i E_{Ti}>M_0$ GeV cut\footnote{Actually to get a large $\sum_i E_{Ti}$ in a perturbative QCD event we have to produce a high $E_T$ jet or to consider the event with a large number of multiple parton interactions. The last possibility is strongly suppressed by the LRG survival factor.} (which we had used further)  we generate the "Hard QCD" events with a minimum transverse momentum of 10 GeV using PYTHIA. Using lower $p_{T}$ samples from PYTHIA, we concluded that the contribution of lower $p_{T}$ events is negligible.

\section{Results}

\begin{figure}[t]
\includegraphics[trim=0.0cm 0cm 0cm 0cm,width=0.5\textwidth]{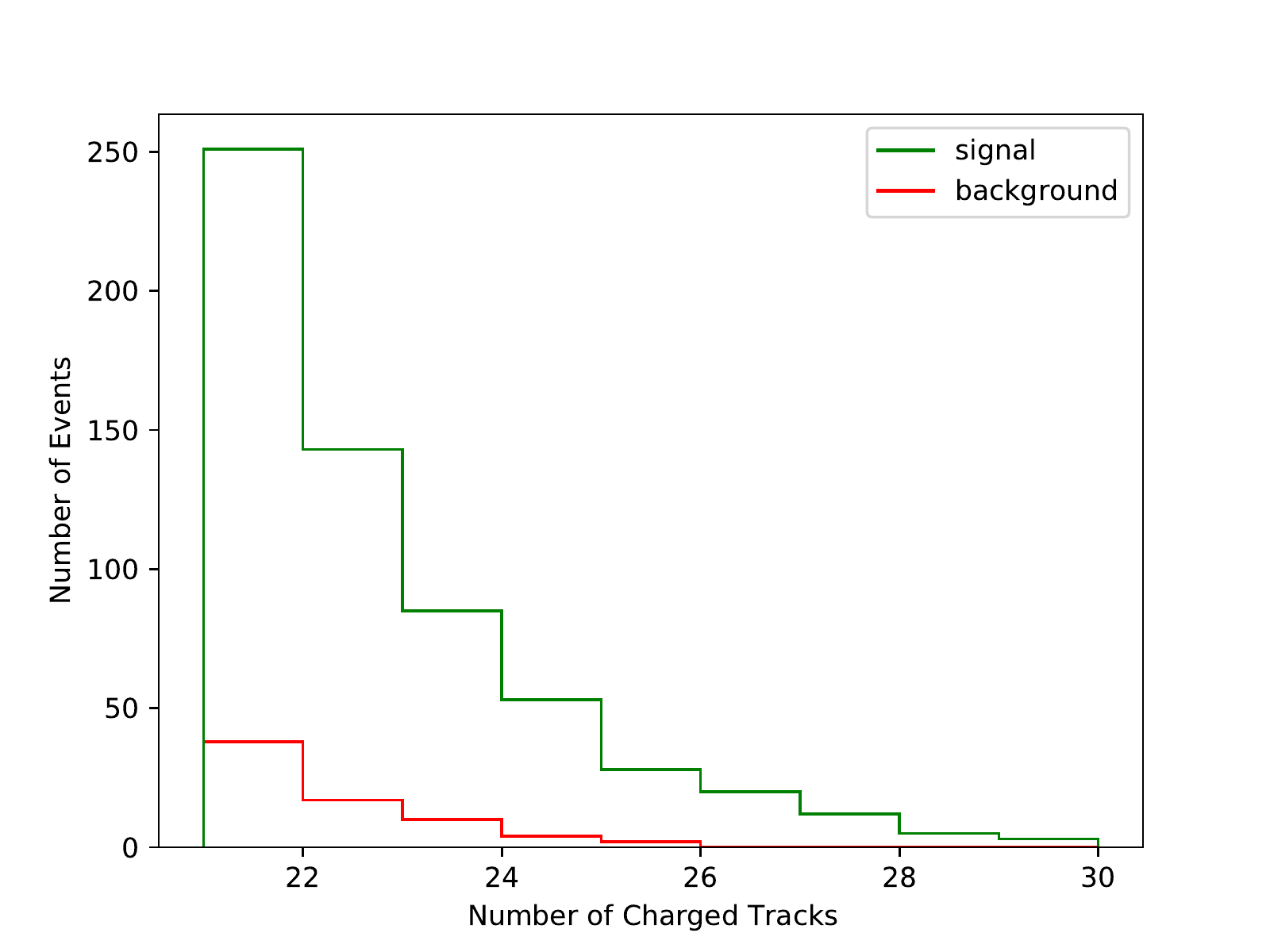}
\includegraphics[trim=0.0cm 0cm 0cm 0cm,width=0.5\textwidth]{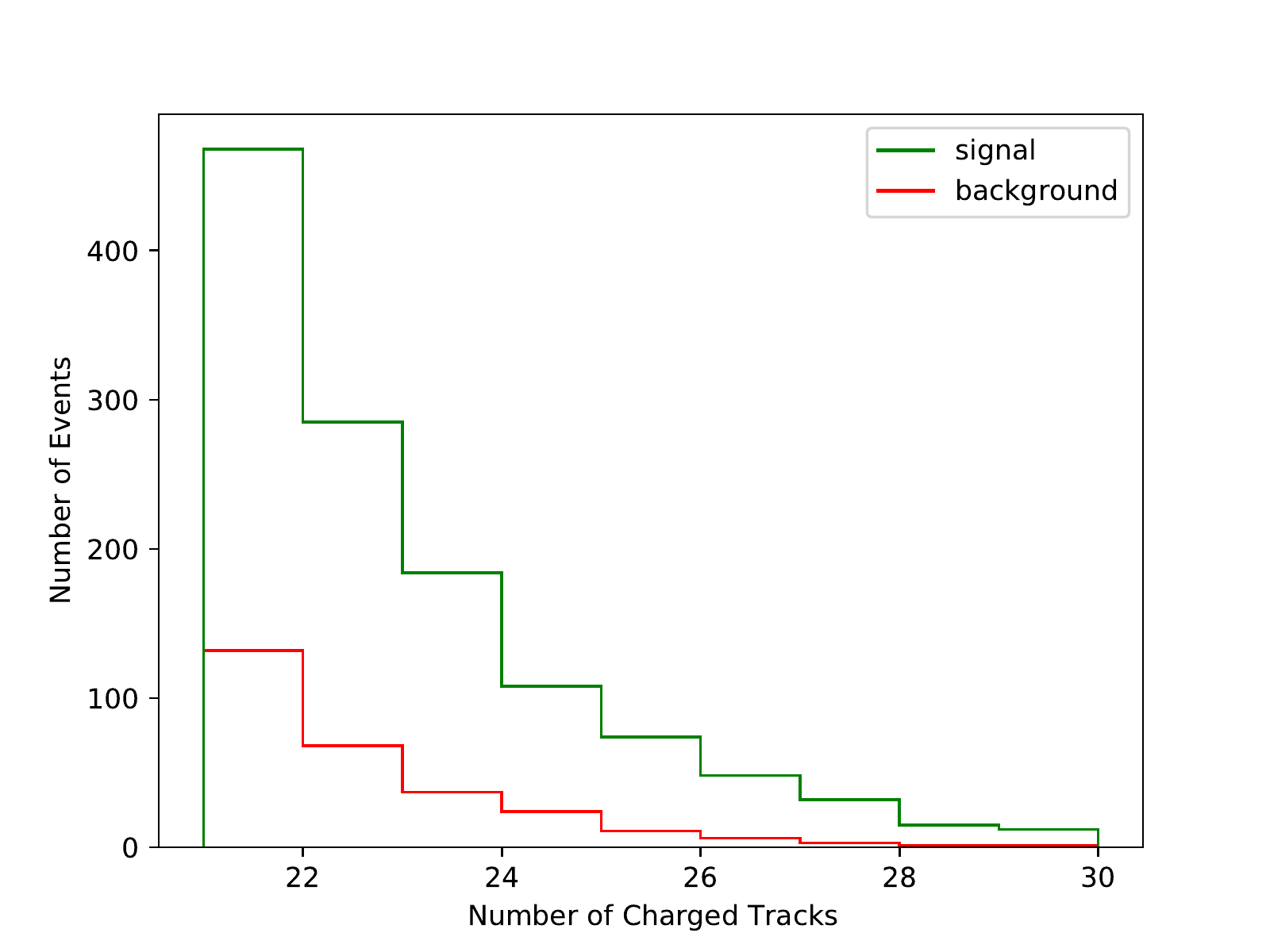}
\caption{\small Multiplicity distribution of charged hadrons produced in the events with the instanton (green) in comparison with the expected background (red).  
The number of events is normalised to the integrated luminosity $L=1\,\mbox{pb}^{-1}$ and $\Delta\ln(x_{Pom})=1$ interval. We required events to have $\sum_i E_{T,i}>15$ GeV and $N_{ch}>20$, summing only over charged particles in the region $0<\eta<2$ with $p_{T}>0.5$ GeV, with an additional constraint  that there is no charged particle in this region with $p_{T}>2$ GeV (left figure), or no charged particle in the region $-2<\eta<2$ with $p_{T}>$ 2.5 GeV (right figure).} 
\label{f3}
\end{figure}

As an example we consider instanton production at 14 TeV in the events with the leading proton momentum fraction $x_L=1-x_{Pom}=0.997$. That is we calculate the cross-section $d\sigma/d\ln(1/x_{Pom})$ at $x_{Pom}=0.003$. This may be considered as the total observed cross-section when/if one integrates over the interval of $x_{Pom}$ from  0.0018 to 0.005 (i.e. $\Delta\ln(x_{pom})$ interval equal~to~1).\footnote{Note that in this region the dependence of single dissociation cross-section  $d\sigma/d\ln(1/x_{Pom})$ is practically flat as is shown in~\cite{SD,ATL,cms}.}

\medskip

\begin{figure} [t]
\includegraphics[trim=0.0cm 0cm 0cm 0cm,width=0.5\textwidth]{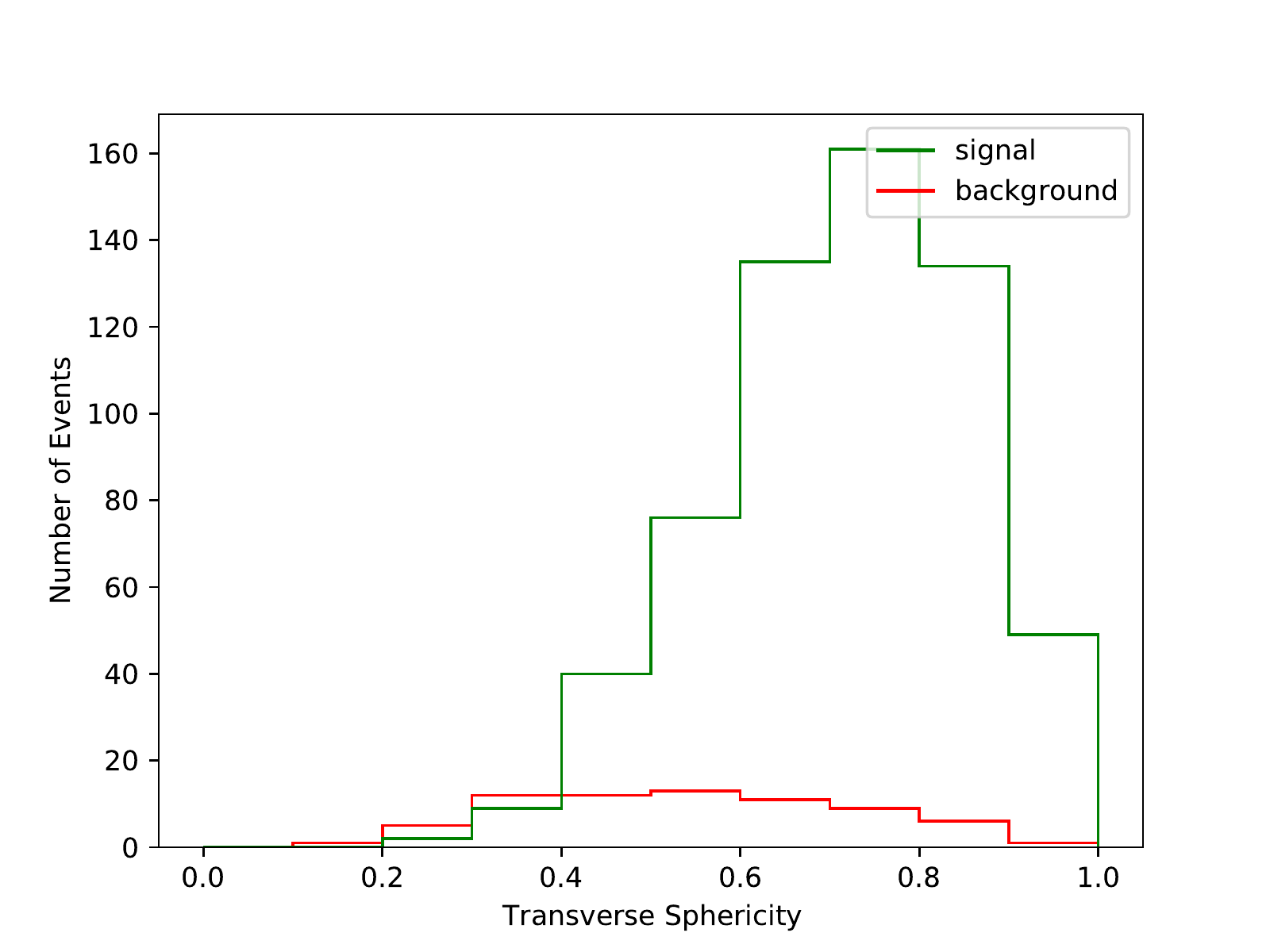}
\includegraphics[trim=0.0cm 0cm 0cm 0cm,width=0.5\textwidth]{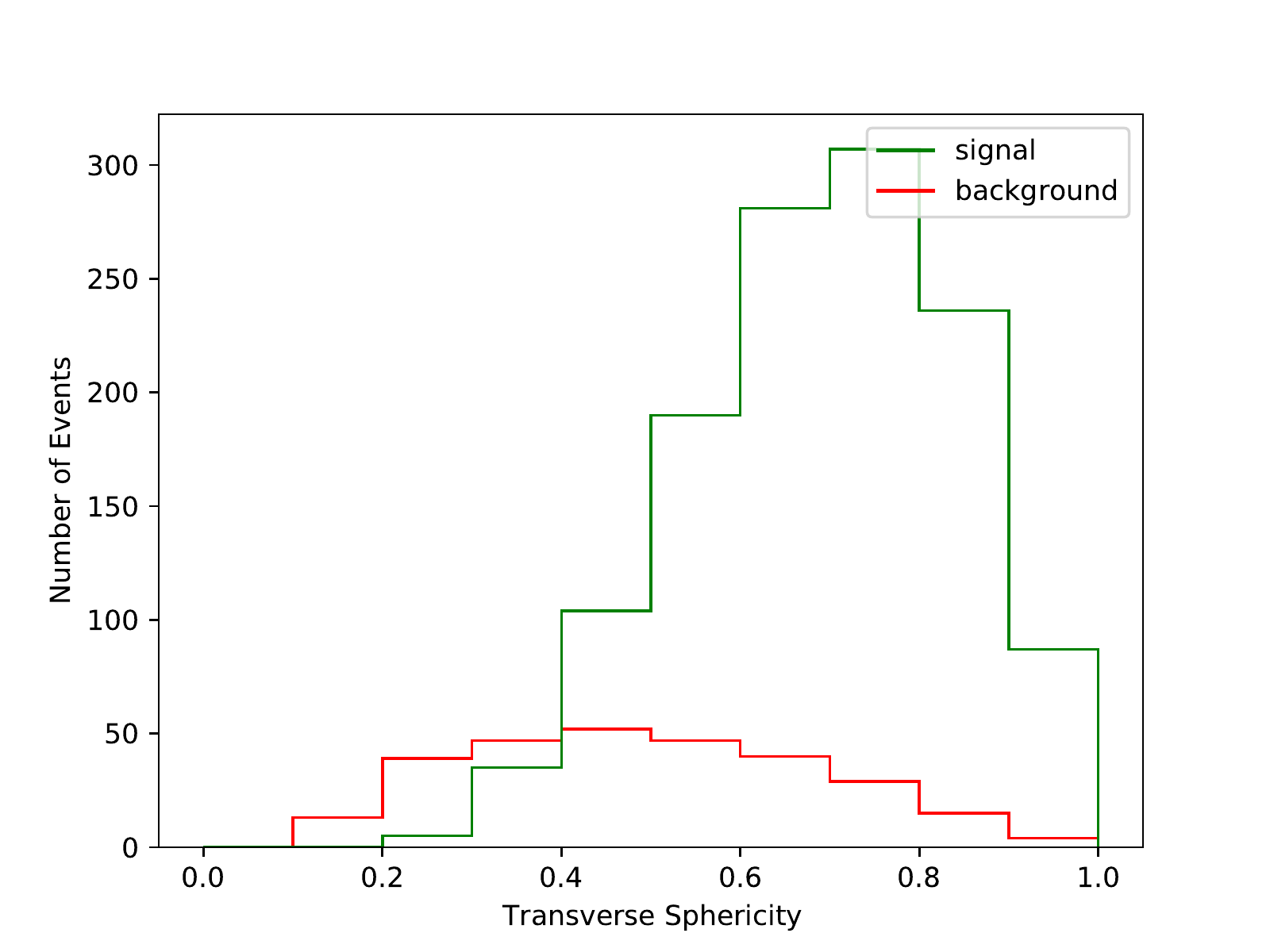}
\caption{\small Distribution over the transverse sphericity $S_T$, Eq.~\eqref{st1}, of the charged hadrons produced in the events with the instanton (green) in comparison with the expected background (red). The selection criteria used are the same as those described in Fig.~\ref{f3}}.
\label{f4}
\end{figure}

Our results are presented in Figs.~\ref{f3} and \ref{f4}. 
To further suppress the background we impose the following set of experimental selection criteria that look realistic for the present ATLAS and CMS detectors.
We
exclude the very low $p_T$ particles and consider only the secondary hadrons with transverse momentum $p_T>0.5$ GeV produced within the $0<\eta<2$ rapidity interval, as shown in 
Fig.~\ref{f2}. We select the events for which the total transverse energy of the charged secondaries (with $p_T>0.5$ GeV and $0<\eta<2$)\footnote{We shifted the $\eta$ interval in the direction opposite to the LRG (or the leading proton which has the negative rapidity) in order not to have 
gluons with too large $x$ (where the gluon density rapidly decreases with increasing $x$)
 from the pomeron PDF. Recall that for $x_{Pom}=0.003$ the energy of the `incoming' pomeron is 21 GeV only.}  is $\sum_i E_{T,i}>15$~GeV and the number of corresponding charged tracks $N_{ch}>20$. We also demand that there are no charged particles in this region with $p_{T}>2$~GeV [left Figs.~\ref{f3} and \ref{f4}]. Plots on the right hand side in Figs.~\ref{f3}-\ref{f4} implement the constraint that no charged particles are present in the region $-2<\eta<2$ with $p_{T}>$ 2.5 GeV.
  
 \medskip

As is seen in Fig.~\ref{f4} after the proposed cuts the instanton signal exceeds the background by a factor of over (left) $8$ (right) $4$ and for $S_T>0.5$ the signal to background ratio S/B is even higher.  Notably, in the $N_{ch}>20$ region the expected cross-section is still rather large ($\sim$~1~nb). This allows one to observe the instanton production in low luminosity runs without  pile-up problems.

\medskip

\begin{figure} [t]
\includegraphics[trim=0.0cm 0cm 0cm 0cm,width=0.5\textwidth]{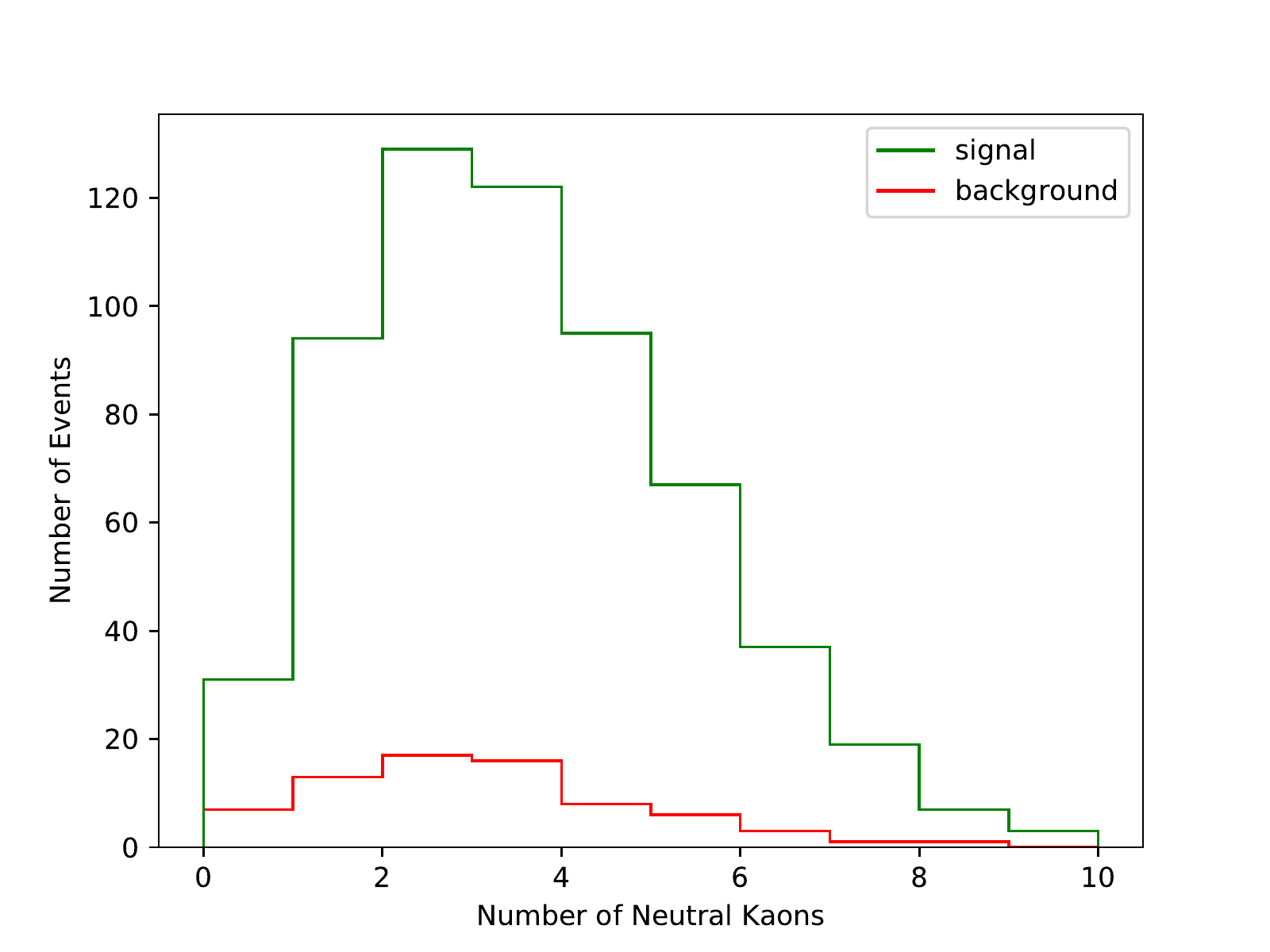}
\includegraphics[trim=0.0cm 0cm 0cm 0cm,width=0.5\textwidth]{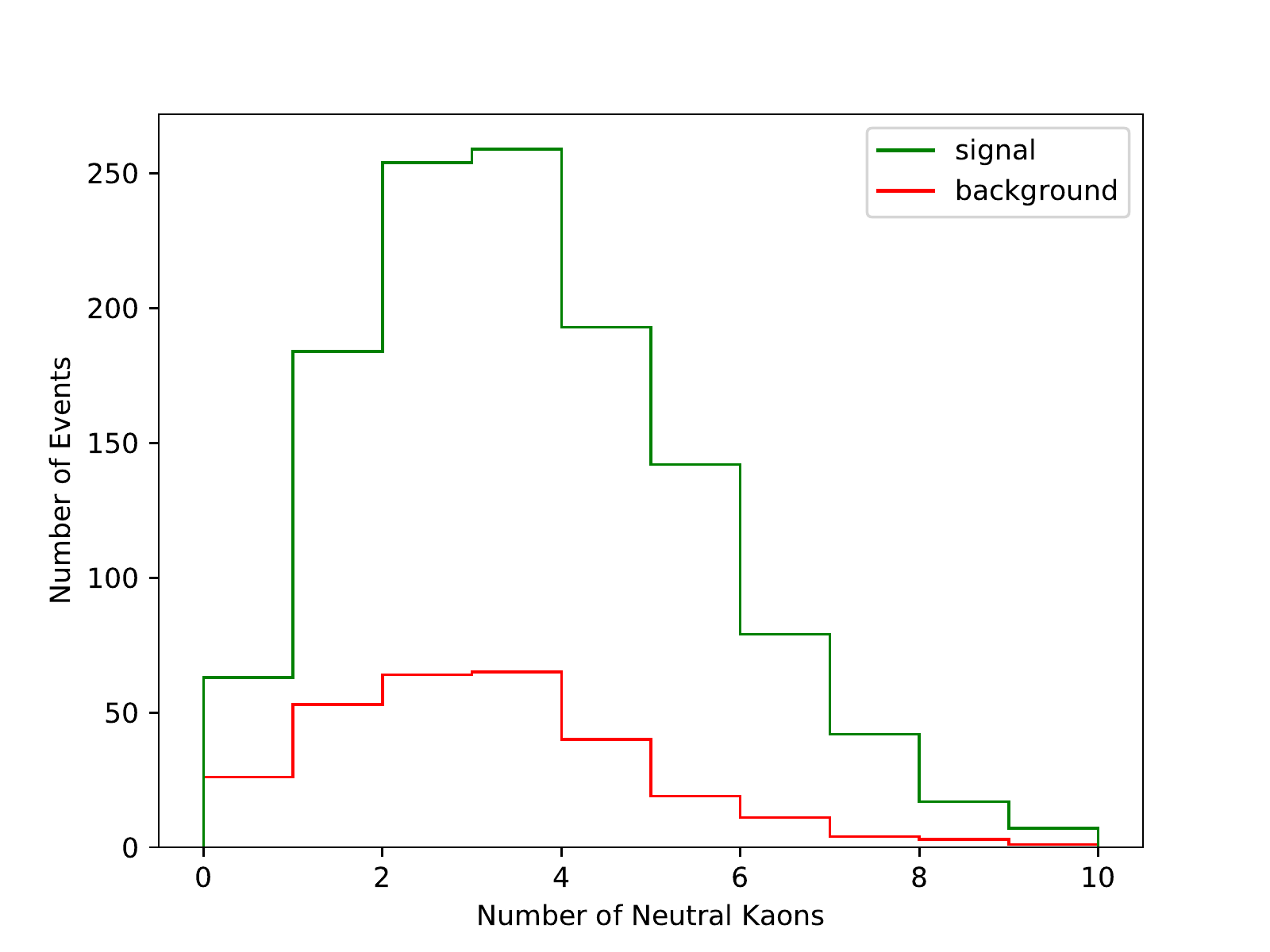}
\caption{\small Distribution over the number of neutral kaons produced in the events with  the instanton (green) in comparison with the expected background (red). 
The selection criteria used are the same as those described in Fig.~\ref{f3}.}
\label{fK}
\end{figure}

The sphericity distribution is shown in Fig.~\ref{f4}. It is clearly seen that while the background sphericity falls for $S_T>0.5$  the instanton signal is mainly concentrated at $S_T\sim0.7-0.8$. For $S_T>0.8$ the expected instanton contribution is a factor of 10 greater than the background (for both sets of selection criteria).

 \medskip
 
 Finally, Fig.~\ref{fK}  shows the distribution over the number of neutral kaons.~\footnote{This can be observed via the  $(K^0)_s\to \pi^+\pi^-$ decay.} The selection criteria used are the same as those described in Fig.~\ref{f3}. As expected the kaon multiplicity is larger in the events containing the instanton
  and for $N_{K^0}>5$ the S/B ratio is about 5 (for both sets of selection criteria).

\section{Discussion}

The possibility of direct experimental observation of QCD instantons at hadron colliders has recently attracted a fair amount of attention~\cite{KKS,KMS,12}. The authors of \cite{KMS} evaluated the sensitivity of the  LHC and Tevatron high- and low-luminosity runs, and \cite{12} obtained a first limit based on existing Minimum Bias data.

\medskip

Extraction of the instanton signal on top of the underlying event at the LHC, not relying on events with large rapidity gaps, is a challenging task. One problem, not considered in \cite{12}, is how to select in the experiment the events corresponding to the chosen $M_{\rm inst}$ interval. It is impossible to distinguish the jet emitted by the instanton from the jet created during the parton showering or 
 due to the presence of a perturbative QCD multi-jet `hedgehog' contribution (the latter is another problem that has not been addressed until now). 
 
 \medskip
 
The idea to observe instantons  in the pomeron-induced process,
namely in the central exclusive production,
\begin{equation}
\label{eq:DPE}
pp~\to~p~+~X~+~p\
\end{equation}
was first put forward in \cite{Shuryak:2003xz} (see also 
\cite{Shuryak:2021iqu}).
This was prompted by the reported  
UA8 observation \cite{Brandt:2002qr} of the large enhancement
in the pomeron-pomeron cross-section at invariant masses $2 < M< 5$ GeV
with roughly isotropic distribution of secondaries.
Unfortunately, we (present authors) do not have theoretical control over such low-mass instantons, or for that matter other non-perturbative contributions in the infrared. In these settings, there is no external mass parameter and it is not clear how to distinguish instanton production from the background or, for example, from the glueball production. On the other hand
searching for a larger-mass instanton in pomeron-pomeron collisions has no advantage in comparison with instanton production in events with one LRG, but in the case of pomeron-pomeron collisions the expected cross-section is much smaller. 

\medskip
\subsection{Other instanton production sub-processes }
\medskip

Up to now we were only concentrating on the 2-gluon initiated process \eqref{e2}, but one should also consider other quark, anti-quark and gluon initiated 2-parton processes, such as,
 \begin{eqnarray}
 g+ u_L &\to &   n_g \times g\,+\, u_R + \sum_{f=1}^{N_f-1} (q_{Rf} +\bar{q}_{Lf})  \,, \label{EQ_i2a} \\
 u_L +\bar{u}_R &\to &  n_g \times g  +\sum_{f=1}^{N_f-1} (q_{Rf} +\bar{q}_{Lf}) \,,  \label{EQ_i2b} \\
u_L + d_L &\to &   n_g \times g\,+\, u_R + d_R+ \sum_{f=1}^{N_f-2} (q_{Rf} +\bar{q}_{Lf})\,,
 \label{EQ_it2c}
\end{eqnarray}
and also include the contributions from multi-parton initial states.

We first consider the 2-parton initiated processes. Note that there is no interference between all these sub-processes since the fermion content of the external states in \eqref{e2},\eqref{EQ_i2a}-\eqref{EQ_it2c} is different. Thus, the total contribution comes from adding the individual cross-sections from these processes, in other words, summing over the integrals \eqref{EQ_i_hadr_gg} with the appropriate choice of pdfs for each of the partons in the initial state. Since each of these contributions is manifestly positive, our already computed contribution from the gluon fusion process \eqref{e2} cannot be reduced. It is also clear that one cannot expect any sufficient enhancement to it from accounting for the additional  
channels \eqref{EQ_i2a}-\eqref{EQ_it2c}. This is because for the parton-level cross-section most of the expression in \eqref{EQ_opt2}, in particular its exponent, remains the same;\footnote{Up to now we have only included the initial-state corrections arising from the high-energy limit of the gluon propagator in the instanton background, but as was shown in~\cite{Khoze:2020paj},
the same result \eqref{EQ_mff}  is expected to hold also for fermions in the initial state.} 
only the four field insertions for the external states vary. 
On the other hand, as we have seen, the dominant contributions to hadronic cross-sections come from the low-$x$ region, and as a result, our gluon fusion instanton sub-process is expected to be the dominant one.

\medskip

Next, we would like to discuss the effects of multi-parton initial states in the instanton production. To do this we would like to distinguish between the 
two cases: one, where multi-partons entering the instanton have originated from just the two parent partons, and the second case, where the multi-parton initial state has a genuine multi-parton origin, with all 3 or 4 partons being emitted from the proton and/or the pomeron. 

The first case, where the multi-partons in the initial state can be traced to just two parent partons is already accounted for by the resummed quantum corrections from the initial states. Such effects are precisely what is included into the initial-initial state and the initial-final state propagators in the instanton background computed in \cite{Mueller:1990qa}. Our approach takes into account the initial-initial (i.e. hard-hard) quantum effects by explicitly including the Mueller form-factor term \eqref{EQ_mff}. The initial-final state (or hard-soft interactions) were not directly included in our approach,\footnote{This is done to avoid double-counting, since the radiative exchanges involving final-state partons are already accounted for by the optical theorem.} however, the form-factor arising from all such quantum corrections is of the same generic exponential form
\cite{Mueller:1990qa},
 \begin{equation}
\exp\Bigl(-\, \frac{\alpha_s(\mu_r)}{16\pi}\,\rho^2 E^2\, \sum_{i,j} a_{ij} \log a_{ij}\Bigr),\quad{\rm where}\quad 
a_{ij} := p_i\cdot p_j /E^2\,,
\label{EQ_mffhs}
\end{equation} 
as the initial-initial state form-factor in \eqref{EQ_mff}. Hence we do not expect a significant difference in the final result from the inclusion of these effects.

The second case corresponds to {\it genuine} multi-parton initial states, and it is accounted for by adding the contribution to the total cross-section from the integrals of the type \eqref{EQ_i_hadr_gg}, now with multiple pdfs, one for each parton in the initial state. Like in the two-particle case, discussed earlier, this is a manifestly positive contribution that cannot destructively interfere with our two-gluon contribution. In summary, we expect that a more careful inclusion of all instanton production sub-processes could only enhance the gluon fusion contribution (and only moderately).

\subsection{Theory uncertainties}
\medskip
Note that all the present calculations should not be considered  as the precise theoretical predictions. There are a number of uncertainties.

First of all, one could expect sizeable corrections to the assumed factorization of instanton and anti-instanton collective coordinate integration measure in the forward elastic scattering amplitude \eqref{EQ_opt2},
which strictly holds only at infinitely large inter-instanton separations $R^2/(\rho\bar{\rho}) \gg 1$. At finite separations, the integration measure was shown  in~\cite{Yung:1987zp} to satisfy,
\begin{equation}
d\mu_{I\bar{I}}\,=\, d\mu_{I}\, d\mu_{\bar{I}}\left(1+{\cal O}\left(\frac{1}{z^2}\right) \right)\,,
\end{equation}
where $z$ is given by \eqref{EQ_defz}. In our calculation the mean value of $R/\rho$ was $\simeq$1.55 at all energies, which corresponds to
$z\simeq 4.16$, or $1/z^2 \simeq 0.058$ which is a reassuringly small value to be optimistic about the validity of the approximate factorization of instanton integration measures (and thus the factorization of the determinants of quadratic fluctuation operators in the instanton--anti-instanton background). 

Hence, in terms of the conformal $z^2$ variable the individual instanton and the anti-instanton are reasonably well-separated.
For this reason, we are similarly confident about the validity of the instanton--anti-instanton action expression~\eqref{EQ_defSz} that was computed for a particular exactly-solvable model~\cite{Yung:1987zp,Khoze:1991mx} of the instanton-anti-instanton valley configuration.
What is more difficult to estimate, however, is the role played 
by the unknown higher-order perturbative corrections to the re-summed quantum exchanges from the initial states in~\eqref{EQ_mff}.

Furthermore, there is a rather large theoretical uncertainty in
evaluation of the gap survival factor $S^2$, which also depends on the details of the particular subprocess (see~\cite{LRG}).
Therefore  we
prefer to use the experimental value measured in diffractive dijet
production\cite{Ddijet,Ddijet2}. We evaluate the corresponding
uncertainty as $\pm 50$\%.
    
 We should also  note a potentially strong dependence of the predicted cross-section on the values of the renormalization, $\mu_R$, and factorization, $\mu_F$, scales. In particular, it was found in~\cite{Mangano:2021udl} that replacing the natural scale $\mu_F=1/\rho$ by the value of $\mu_F=M_{\rm inst}$ enhances the instanton production cross-section for instantons of mass larger than $M_{\rm inst}=50$ GeV computed in \cite{KKS}  by approximately a factor of five. 
  
Besides this, as we have already noted, the present calculations  are based on the simplest $gg\to Instanton$ subprocess neglecting the possibility to produce the instanton in collision with quarks or with a pair of gluons, like $g+{gg}\to Instanton$ and so on. Based on our earlier discussion in Section~7.1, we consider it unlikely that these new subprocesses would change the qualitative features of the expected signal, but they may enlarge the final cross-section.  
 
\section{Conclusions}

We propose to search for the QCD instantons at the LHC in events with Large Rapidity Gaps and in this paper we carried out the first preliminary investigation of this search strategy. These LRG events are selected either by detecting the forward leading proton with $x_L$ very close to 1 or by observing no hadron activity in the forward calorimeter. We discussed the main sources of background for the low and the high mass instantons and showed that background
 for the low mass instantons, which comes mainly from multiple parton interactions, is effectively suppressed in the LRG events by the small gap survival factor $S^2\leq 0.1$. This allowed us to find the kinematical domain where the signal from $M_{\rm inst}\sim 20-60$ GeV instantons reliably exceeds background. 
 
 \medskip
 Even with these rather strong cuts in place, the expected instanton cross-section remains sufficiently 
 large ($\sim $ 1 nb) to effectively produce and probe QCD instantons at the LHC, at low luminosity runs, avoiding pile-up problems.

\section*{Acknowledgments}

We thank Valery Schegelsky and Michael Spannowsky for useful discussions. VVK is supported by the STFC under grant ST/P001246/1. DLM is supported by an STFC studentship.


\end{document}